\documentclass[11pt]{article} 

\usepackage{amsfonts}
\usepackage{amssymb}
\usepackage{amsmath}
\usepackage{graphicx}

\newcommand{\ket}[1]{|#1\rangle}

\newcommand{\bra}[1]{\langle #1|}

\newcommand{\bracket}[2]{\langle #1|#2\rangle}

\newcommand{\op}[1]{\operatorname{#1}}
\def \qed {\hfill \rule{0.2cm}{0.2cm}\vspace{3mm}}
\newenvironment{mylist}[1]
	{\begin{list}{}{\setlength{\leftmargin}{#1}
	\setlength{\rightmargin}{0.0cm}\setlength{\labelsep}{1.3mm}
	\setlength{\labelwidth}{0.8cm}\setlength{\itemsep}{0.2cm}}}
	{\end{list}}

\newtheorem{theorem}{Theorem}[section]
\newtheorem{lemma}[theorem]{Lemma}

\newtheorem{definition}{Definition}[section]
\newenvironment{defn}{\begin{definition}\em}{\end{definition}}

\begin{document}

\newlength{\twidth}

\title{On quantum and classical space-bounded processes with\\
algebraic transition amplitudes}

\author{John Watrous\thanks{
	Department of Computer Science, University of
	Calgary, Calgary, Alberta, Canada T2N 1N4.  Email:
	{\tt jwatrous@cpsc.ucalgary.ca}.  This research was supported by
	Canada's {\sc Nserc} and was done while the author was at the
	D\'epartement IRO, Universit\'e de Montr\'eal, Montr\'eal, Qu\'ebec,
	Canada H3C 3J7.}}

\maketitle

\thispagestyle{empty}

\begin{abstract}
We define a class of stochastic processes based on evolutions and measurements
of quantum systems, and consider the complexity of predicting their long-term
behavior.
It is shown that a very general class of decision problems regarding these
stochastic processes can be efficiently solved classically in the space-bounded
case.
The following corollaries are implied by our main result for any
space-constructible space bound $s$ satisfying $s(n) = \Omega(\log n)$.
\begin{mylist}{3mm}

\item[$\bullet$]
Any space $O(s)$ uniform family of quantum circuits acting on $s$ qubits and
consisting of unitary gates and measurement gates defined in a typical way by
matrices of algebraic numbers can be simulated by an unbounded error space
$O(s)$ ordinary (i.e., fair-coin flipping) probabilistic Turing machine, and
hence by space $O(s)$ uniform classical (deterministic) circuits of depth
$O(s^2)$ and size $2^{O(s)}$.
The quantum circuits are not required to operate with bounded error and may
have depth exponential in $s$.

\item[$\bullet$]
Any (unbounded error) quantum Turing machine running in space $s$, having
arbitrary algebraic transition amplitudes, allowing unrestricted measurements
during its computation, and having no restrictions on running time can be
simulated by an unbounded error space $O(s)$ ordinary probabilistic Turing
machine, and hence deterministically in space $O(s^2)$.
\end{mylist}
We also obtain the following classical result:
\begin{mylist}{\parindent}
\item[$\bullet$]
Any unbounded error probabilistic Turing machine running in space $s$ that
allows algebraic probabilities and algebraic cut-point can be simulated by a
space $O(s)$ ordinary probabilistic Turing machine with cut-point 1/2.
\end{mylist}

Our technique for handling algebraic numbers in the above simulations may be
of independent interest.
It is shown that any real algebraic number can be accurately approximated by a
ratio of GapL functions.
\end{abstract}


\section{Introduction}
\label{sec:introduction}

In this paper we investigate the complexity of predicting the long-term
behavior of stochastic processes induced by evolutions and measurements of
discrete quantum mechanical systems.
The processes considered, which we call {\em selective quantum processes},
describe the classical outputs obtained when operations called {\em selective
quantum operations} are iterated on finite state quantum systems.
Quantum Turing machine and quantum circuit computations may be viewed as
specific examples of such processes.
Selective quantum operations are quite general and include unitary evolutions
and positive operator valued measures (POVMs).
The main result proved in this paper regards the space required for classical
machines to solve decision problems based on selective quantum processes; we
postpone the formal statement of this result until after necessary background
material has been discussed.

Although quantum computation offers the potential for exponential speed-up over
classical computation for certain problems, such as integer factoring and
discrete logarithms \cite{Shor97}, the analogous situation does not hold with
regard to the amount of space required by quantum machines vs.~classical
machines.
It was proved in \cite{Watrous99} that quantum Turing machines satisfying
certain restrictions (discussed below) and running in a given space bound $s$
(for $s$ space-constructible and satisfying $s(n) = \Omega(\log n)$) can be
simulated by space $O(s)$ probabilistic Turing machines in the unbounded error
setting.
The assumptions made on the quantum machines were that only rational transition
amplitudes could be used, and only a very limited class of observations during
the computations were permitted.
By applying our main result to quantum Turing machine computations, we
extend the above result to machines allowing algebraic transition amplitudes
rather than just rational ones, and further to machines allowing unrestricted
measurements during their computations.
Our main result also has implications for bounded-width quantum circuits:
any space $O(s)$ uniform family of quantum circuit acting on $s$ qubits that
consist of unitary gates and measurement gates defined by matrices of
algebraic numbers (as described later in Section~\ref{sec:main_result}) can be
simulated by an unbounded error space $O(s)$ probabilistic Turing machine.

A well-known result of Borodin, Cook, and Pippenger \cite{BorodinC+83} states
that any unbounded-error space $s$ probabilistic Turing machine computation
can be simulated deterministically by uniformly generated depth $O(s^2)$
circuits with size polynomial in $2^s$, and hence in deterministic space
$O(s^2)$ (i.e.,
$\mathrm{PrSPACE}(s)\subseteq\mathrm{NC}^2(2^s)\subseteq\mathrm{DSPACE}(s^2)$).
Thus, our results imply the existence of efficient parallel (classical)
algorithms (in the sense that NC represents a class of efficiently
parallelizable problems) for predicting the long-term behavior of quantum
systems of modest size.
It is interesting to note that the complexity of the algorithm implied by our
technique is independent of the running time of the quantum process: in
(parallel) time proportional to $s^2$, we may predict the behavior of a
quantum system consisting of $s$ components (e.g., qubits) that runs for an
arbitrary number of steps.

The technique we use to prove our main result is similar to one used in
\cite{Watrous99}, and has previously been used to prove results in classical
space-bounded computation (for instance in
\cite{AllenderO96,BorodinC+83,Jung85}).
Essentially, the technique is to manipulate matrices that govern the
stochastic processes being considered in order to predict their long-term
behavior (rather than simulating the processes directly), with the matrix
manipulations being performed in a very space-efficient manner.
In the present case, we must modify this technique in order to handle
algebraic matrices rather than rational ones, and we must reformulate the
quantum processes we are considering in the framework covered by the technique.
Section~\ref{sec:proof_of_main} describes this in detail.


\section{Quantum processes}
\label{sec:quantum_processes}

In this section we briefly review certain facts from quantum computation and
state the definition of selective quantum processes that will be used
throughout this paper.
For a more thorough treatment of quantum computing, we refer the reader to the
surveys of Berthiaume \cite{Berthiaume97} and Kitaev \cite{Kitaev97}, and to
the references therein.
Our definition of selective quantum operations is implicit in
\cite{BrussD+98,NielsenC97}.
A number of the claims made in this section have straightforward proofs
using matrix analysis (see, e.g., Horn and Johnson \cite{HornJ85}), which
we omit.

We restrict our attention to quantum systems having finite classical state
sets; for a given system, we generally denote the classical state set by $S$.
For example, in the case of quantum circuits the set $S$ may be the set of all
0-1 assignments to the wires at some particular level in the circuit, while
in the case of quantum Turing machines $S$ may be the set of all configurations
of the machine subject to some given space bound.
Given a quantum system with fixed classical state set $S$, a {\em pure state}
(or {\em superposition}) of the system is unit vector in the Hilbert space
$\ell_2(S)$.
We use the Dirac notation to represent elements of $\ell_2(S)$;
for each $s\in S$, $\ket{s}$ represents the unit vector corresponding to
the map that takes $s$ to 1 and each $s^{\prime}\not=s$ to 0.
Elements of $\ell_2(S)$ are generally denoted $\ket{\psi}$, $\ket{\phi}$,
etc., and may be specified by linear combinations of elements in the
orthonormal basis $\{\ket{s}:s\in S\}$.
Corresponding to each $\ket{\psi}$ is a linear functional $\bra{\psi}$ that
maps each vector $\ket{\phi}$ to the inner product $\bracket{\psi}{\phi}$
(conjugate-linear in the first argument).

A {\em mixed state} is a state that may be described by a distribution on (not
necessarily orthogonal) pure states.
Intuitively, a mixed state represents the quantum state of a system given that
we have limited knowledge of this state.
A collection $\{\left(p_k,\ket{\psi_k}\right)\}$ such that
$0\leq p_k$, $\sum_k p_k=1$, and each $\ket{\psi_k}$ is a pure state is called
a {\em mixture}: for each $k$, the system is in superposition $\ket{\psi_k}$
with probability $p_k$.
It is the case that different mixtures may yield identical states, in the
sense that no measurement can distinguish the mixtures even in a statistical
sense.
For a given mixture $\{\left(p_k,\ket{\psi_k}\right)\}$, we associate an
$|S|\times |S|$ {\em density matrix} $\rho$ having operator representation
$\rho = \sum_k p_k\ket{\psi_k}\bra{\psi_k}$.
Two mixtures yield different density matrices if and only if there exists a
measurement that can statistically distinguish the two mixtures, and so we
interpret a given density matrix $\rho$ as being a canonical representation of
a given mixed state.
Necessary and sufficient conditions for a given $|S|\times |S|$ matrix $\rho$
to be a density matrix (i.e., represent some mixed state) are (i)~$\rho$ must
be positive semidefinite, and (ii)~$\rho$ must have unit trace.

A {\em selective quantum operation} is a probabilistic mapping that takes as
input a density matrix $\rho$ and outputs a collection of pairs
$(i,\rho^{(i)})$, each with some probability $p_i$; each $\rho^{(i)}$ is a
density matrix and $i$ is a classical output that we take to be an integer for
simplicity.
The output $i$ may be the result of some measurement, although this is not the
most general situation (for example, the system may be measured and part of
outcome may be discarded).
A selective quantum operation $\mathcal{E}$ must be described by a collection
$\{A_{i,j}\,|\,0\leq i\leq m,\,1\leq j\leq l\}$ of $|S|\times|S|$ matrices
satisfying the constraint
$\sum_{i=0}^m\sum_{j=1}^l A_{i,j}^{\dagger}A_{i,j}=I$.
Given such a collection of matrices, we define a function
$p_i:\mathbb{C}^{n\times n}\rightarrow[0,1]$ and a partial function
$E_i:\mathbb{C}^{n\times n}\rightarrow\mathbb{C}^{n\times n}$ as follows:
\begin{eqnarray*}
p_i(\rho) & = &\op{tr}\left(\sum_{j=1}^l A_{i,j}\rho A_{i,j}^{\dagger}\right)\\
E_i(\rho) & = & \frac{1}{p_i(\rho)}\sum_{j=1}^l A_{i,j}\rho A_{i,j}^{\dagger}.
\end{eqnarray*}
(In case $p_i(\rho) = 0$, $E_i(\rho)$ is undefined.)
Now, on input $\rho$, the output of $\mathcal{E}$ is defined to be
$(i,E_i(\rho))$ with probability $p_i(\rho)$ for each $i$.
It may be verified that for any density matrix $\rho$, we have
$0\leq p_i(\rho)$ and $\sum_i p_i(\rho)=1$, and furthermore that each
$E_i(\rho)$ is a density matrix (and so the above definition is sensible).
We also define functions $F_0,\ldots,F_m$ as
$F_i(\rho)=\sum_{j=1}^l A_{i,j}\rho A_{i,j}^{\dagger}$.
It will simplify matters when calculating unconditional probabilities to
consider these functions.

Finally, a {\em selective quantum process} is a stochastic process
$\{R_t\,|\,t\in\mathbb{N}\}$, where each $R_t$ is a random variable whose
value corresponds to the classical output of a selective quantum operation.
A selective quantum operation
$\mathcal{E} = \{A_{i,j}\,|\,0\leq i\leq m,\,1\leq j\leq l\}$ and an
initial density matrix $\rho_{init}$ induce a selective quantum process
$\{R_t\,|\,t\in\mathbb{N}\}$ as follows: for $n\geq 1$ and
$r_1,\ldots,r_n\in\{0,\ldots,m\}$, the probability that $R_1,\ldots,R_n$ take
values $r_1,\ldots,r_n$ is the probability that, if the selective quantum
operation $\mathcal{E}$ is iterated $n$ times given initial state
$\rho_{init}$, the resulting classical outputs will be $r_1,\ldots,r_n$.
Thus, the probability that $R_n$ takes a particular value $r_n$ depends on the
values taken by $R_1,\ldots,R_{n-1}$ in the following way:
\[
\op{Pr}\left[R_n = r_n|R_1 = r_1,\ldots, R_{n-1} = r_{n-1}\right]\:
= \: p_{r_n}\left(E_{r_{n-1}}\circ\cdots\circ E_{r_1}(\rho_{init})\right).
\]
It may be proved by induction that for a selective quantum process
$\{R_t\,|\,t\in\mathbb{N}\}$ induced by $\mathcal{E}$ and $\rho_{init}$, we
have $\op{Pr}[R_1=r_1,\ldots,R_n=r_n]=
\op{tr}(F_{r_n}\circ\cdots\circ F_{r_1}(\rho_{init}))$
for all $n\geq 1$.


\section{GapL functions and PL}
\label{sec:background_GapL}

Next, we recall some definitions and facts from counting complexity and
space-bounded complexity.
Counting complexity is a powerful technique that has its origins in the work
of Valiant \cite{Valiant79}, and has had a number of applications in
complexity theory (including in quantum computing
\cite{FennerG+98b,FortnowR99}).
For further information on counting complexity, see the survey of Fortnow
\cite{Fortnow97} and the references therein.
Counting complexity was applied to space-bounded computation in
\cite{AllenderO96,AlvarezJ93}, to which the reader is referred to for proofs
of the theorems stated in this section.
For more general background information on space-bounded computation, see
Saks \cite{Saks96}.

Consider a nondeterministic Turing machine $M$ running in logspace.
On each input $x$ there are some number of computation paths that lead to
an accepting configuration and some number of paths that lead to a rejecting
configuration; we denote these numbers by $\#M(x)$ and $\#\overline{M}(x)$,
respectively.
\begin{definition}
A function $f:\Sigma^{\ast}\rightarrow\mathbb{Z}$ is a GapL function
($f\in\mathrm{GapL}$) if there exists a logspace nondeterministic Turing
machine $M_f$ such that $f(x) = \#M_f(x) - \#\overline{M_f}(x)$ for every
input $x$.
\end{definition}

GapL functions characterize the class PL as follows:
\begin{theorem}
\label{thm:PL_GapL}
Let $A\subseteq\Sigma^{\ast}$.
Then $A\in\mathrm{PL}$ if and only if there exists $f\in\mathrm{GapL}$ such
that $x\in A\:\Leftrightarrow\:f(x)>0$ for every $x\in\Sigma^{\ast}$.
\end{theorem}

Any integer function computable in logspace is necessarily a
GapL function: $\mathrm{FL}\subseteq\mathrm{GapL}$.
GapL functions are quite useful for proving results in space-bounded complexity
due to the fact that various closure properties of GapL functions hold:
\begin{theorem}
\label{thm:GapL_closure}
Let $f\in\mathrm{GapL}$, let $g\in\mathrm{FL}$, and let $p$ be an integer
polynomial.
Define functions $h_1$, $h_2$, and $h_3$ as follows: $h_1(x)=f(g(x))$,
$h_2(x)=\sum_{i = 0}^{p(|x|)}f(x,i)$, and $h_3(x)=\prod_{i=0}^{p(|x|)}f(x,i)$.
Then $h_1$, $h_2$, and $h_3$ are in GapL.
\end{theorem}
Here it is assumed that nonnegative integers are identified with their
encodings as binary strings in the usual way.

There is a close relationship between GapL functions and the determinant
function, which is crucial for the proof of our main theorem:

\begin{theorem}
\label{thm:determinant}
Let $f\in\mathrm{GapL}$ and let $p$ be any integer polynomial satisfying
$p(n)\geq 1$ for $n\geq 0$.
For each $x\in\Sigma^{\ast}$ let $A(x)$ be a $p(|x|)\times p(|x|)$ matrix
defined as $A(x)[i,j] = f(x,i,j)$, and define $g(x) = \op{det}(A(x))$.
Then $g\in\mathrm{GapL}$.
\end{theorem}
See \cite{AllenderA+97} for a proof of this theorem.


\section{Main result and applications}
\label{sec:main_result}

Now we are prepared to state the main theorem.

\begin{theorem}
\label{thm:main}
Let $p$ and $q$ be integer polynomials satisfying $p(n),q(n)\geq 1$ for
$n\geq 0$, let\linebreak
$\{\alpha_1,\ldots,\alpha_k,\beta\}$ be a finite collection of
algebraic numbers with $\beta$ real and in the range $[0,1]$, and let $a_l$,
$b_l$, $c_l$, and $d_l$ (for $1\leq l\leq k$) be GapL functions such that each
$b_l$ and $d_l$ is nonzero on all inputs.
For each $x\in\Sigma^{\ast}$, let $\mathcal{E}_x$ be a selective quantum
operation described by $p(|x|)\times p(|x|)$ matrices
$\{A_{x,i,j}\,|\,0\leq i\leq q(|x|),\,1\leq j\leq q(|x|)\}$ defined as follows:
\[
A_{x,i,j}[i^{\prime},j^{\prime}] \:=\: \sum_{l=1}^k \frac{a_l(x,i,j,
i^{\prime},j^{\prime})}{b_l(x,i,j,i^{\prime},j^{\prime})}\,\alpha_l,
\]
and let $\rho_x$ be a $p(|x|)\times p(|x|)$ density matrix specified as
follows:
\[
\rho_x[i^{\prime},j^{\prime}] \: = \: \sum_{l=1}^k
\frac{c_l(x,i^{\prime},j^{\prime})}{d_l(x,i^{\prime},j^{\prime})}\,\alpha_l.
\]
Let $\{R_{x,t}\,|\,t\in\mathbb{N}\}$ be the selective quantum process induced
by $\mathcal{E}_x$ and $\rho_x$ for each $x$.
Then the language
\[
\{x\,|\,\op{Pr}[\exists\,t:\,R_{x,1}=\cdots=R_{x,t-1}=0,\,R_{x,t}=1]>\beta\}
\]
is in $\mathrm{PL}$.
\end{theorem}

\noindent
This theorem is sufficiently general to imply that a wide variety of ``logspace
quantum computations'' can be simulated in PL, including logspace quantum
Turing machine and logarithmic-width quantum circuit computations that have
algebraic transition amplitudes and allow measurements during their
computations.
We now discuss these special cases in more detail.
We conclude this section with a brief discussion on how these results can be
extended to a more general class of space-bounds.

We will begin with quantum circuits, since it will be straightforward to
translate results in this setting to the quantum Turing machine model.
We use the quantum circuit formalization given in \cite{AharonovK+98}, which
allows a very general class of quantum gates including unitary gates and
``measurement gates''.
In order to fix our notation, we first briefly review this formalism.

A $k$ qubit {\em quantum gate} is a linear mapping on the space of
$2^k\times 2^k$ complex matrices induced by a collection of $2^k\times 2^k$
matrices $\{A_1,\ldots,A_m\}$ in a similar manner to selective quantum
operations (but with no classical output).
Specifically, we require $\sum_j A_j^{\dagger}A_j = I$, and we have that this
collection induces the mapping $\rho\mapsto\sum_j A_j\rho A_j^{\dagger}$
on density matrices.
A $k$ qubit gate acts on an ordered $k$-tuple of qubits in the natural way:
if we associate the initial mixed state of the qubits with a $2^k\times 2^k$
density matrix $\rho$, the effect of the gate is given by the above mapping.
An $n$ qubit {\em quantum circuit} is a sequence of quantum gates applied to
ordered subsets of a collection of $n$ qubits.
The action of a $k$ qubit gate on a given ordered subset of $n$ qubits, for
$n>k$, is given by taking the Kronecker product of each matrix $A_j$ with
the $2^{n-k}\times 2^{n-k}$ identity matrix, permuting rows and columns
of the resulting matrices according to which qubits are acted on by the gate,
and applying the resulting operation to the state of the $n$ qubits.

We now define what we mean by a logspace-uniform family of quantum circuits
acting on a logarithmic number of qubits.
Let $\mathcal{G}=\{G_1,\ldots,G_k\}$ be a fixed, finite collection of quantum
gates, each acting on a constant number of qubits and specified by a
collection of matrices as above.
We require that each such matrix have entries that are algebraic numbers.
Let $s$ be a space-constructible function satisfying $s(n) = \Theta(\log n)$.
Since $s$ is sub-linear, we assume the particular input $x$ is not given as
input to the quantum circuit, but rather is input to the deterministic
procedure that generates the circuits.
The input to the circuit is assumed to be $s(|x|)$ qubits each in the
$\ket{0}$ state.
Let $f$ be a mapping that takes an input $x$ and an integer in the range
$\{1,\ldots,r(|x|)\}$ as input, for some polynomial $r$, and outputs an
index of an element in $\mathcal{G}$ along with an ordered subset of
$\{1,\ldots,s(|x|)\}$, specifying a gate and the qubits upon which that gate
is to act.
The quantum circuit generated by $f$ is the sequence of gates given by
$f(x,1),\ldots,f(x,r(|x|))$, applied in order.
One of the qubits is specified as the output of the circuit, and is assumed
to be observed in the 0-1 basis after the circuit has been applied, yielding
acceptance or rejection.
(In case we wish to consider the output of the circuit to be a function,
multiple qubits may be specified as output qubits.)
If the function $f$ can be computed in logspace by a deterministic
Turing machine, then we say the resulting family of circuits is
logspace-uniform.

Now, we claim that for any logspace-uniform family of quantum circuits acting
on a logarithmic number of qubits, the language consisting of those inputs $x$
accepted with probability exceeding a given algebraic cut-point $\beta$ is in
PL, following from Theorem~\ref{thm:main}.
This may be shown by defining a selective quantum operation $\mathcal{E}_x$
acting on $s(|x|)+\lceil\log_2 r(|x|)\rceil$ qubits: the first $s(|x|)$ qubits
represent the qubits in the quantum circuit, and the remaining qubits index
the gates in the circuit.
The selective quantum operation $\mathcal{E}_x$ effectively measures the
qubits indexing the particular gate to be applied, applies that gate
appropriately to the first $s$ qubits, increments the index qubits modulo
$r(|x|)$, and outputs one of the classical results 0 (the computation has not
yet completed), 1 (the circuit accepts), or 2 (the circuit rejects).
Given that the mapping $f$ is logspace computable, it is possible to define
GapL functions $a_l$ and $b_l$ (in fact, each $b_l$ may be constant, taking
value 1, and each $a_l$ may be an FL function) such that $\mathcal{E}_x$ is
given by $p(|x|)\times p(|x|)$ matrices
$\{A_{x,i,j}\,|\,0\leq i\leq q(|x|),\, 1\leq j\leq q(|x|)\}$ for polynomials
$p$ and $q$ as in the statement of Theorem~\ref{thm:main}.
(Since the output of $\mathcal{E}_x$ is always 0,1, or 2, we will have
$A_{x,i,j}=0$ for $i>2$.)
The probability that the circuit accepts is precisely
$\op{Pr}[\exists\,t:\,R_{x,1}=\cdots=R_{x,t-1}=0,\,R_{x,t}=1]$, for
$\{R_{x,1},R_{x,2},\ldots\}$ the process induced by $\mathcal{E}_x$ (along
with the initial density matrix $\rho_x$ describing the initial zero state of
the qubits and initial state of the index qubits---easily seen to be computable
in the sense of Theorem~\ref{thm:main}).
A more formal presentation of this construction will appear in the final
version of this paper.

Next we discuss quantum Turing machine computations.
We are not aware of any systematic treatment of quantum Turing machines that
may perform unitary operations and measurements during their computations, nor
will we attempt to provide such a treatment here.
However, we claim that any reasonable notion of a quantum Turing machine $M$
running in logspace that allows measurements during its computation may be
formulated in terms of a selective quantum process in such a way that the
following holds: the language consisting of all strings accepted by $M$ with
probability exceeding some algebraic cut-point $\beta$ reduces to the language
in the statement of Theorem~\ref{thm:main}, and hence is contained in PL.

For instance, consider a Turing machine consisting of two parts: a classical
part and a quantum part.
The input tape may be considered to belong to the classical part, along with
a classical work tape and a classical portion of the internal state, while
the quantum part consists of a quantum work tape and a quantum portion of the
internal state.
For each local description of the classical part of the machine, we may have a
quantum transition function that specifies the evolution of the quantum part
of the machine in the usual manner (e.g., as described by Bernstein and
Vazirani \cite{BernsteinV97}).
Such ``quantum steps'' may be alternated with ``classical steps'', in which
the classical part of the machine evolves classically, perhaps involving
measurements of the quantum portion of the internal state.
Under the assumption that the specifications of the quantum transition
functions and the measurements of the quantum portion of the internal state
are described by finite collections of algebraic numbers, the problem of
determining if such a machine accepts with probability exceeding some cut-point
$\beta$ reduces to the problem in Theorem~\ref{thm:main}.
This may be argued by referring to the previous discussion on quantum circuits.
Given such a quantum Turing machine, we may represent the work tapes, internal
state, and input tape head position of this machine by the qubits in a
logspace uniform quantum circuit, with the read-only input of the Turing
machine corresponding to the input to the logspace function generating the
circuit.
Similar to the quantum circuit simulation of quantum Turing machines due to
Yao \cite{Yao93}, we may define a quantum circuit that simulates one step in
the Turing machine computation.
As above, we may then define a selective quantum operation $\mathcal{E}_x$
that simulates the action of this circuit, and produces classical output 0, 1,
or 2 as above, with output 1 or 2 corresponding to the situation that the
Turing machine has entered an accepting or rejecting state, respectively.
Here we take advantage of the fact that Theorem~\ref{thm:main} places no
restriction on the running time of the quantum process, since the Turing
machine being simulated need not necessarily halt absolutely or even with
probability 1.
Again we postpone the formal presentation of this construction to the final
version of this paper.

We also note that probabilistic Turing machines having algebraic probability
transitions and algebraic cut-point are a restricted case of the quantum
Turing machines we have considered.
Theorem~\ref{thm:main} thus implies that even in the unbounded error
setting, logspace probabilistic Turing machines having algebraic transitions
and cut-points are equivalent in power to ordinary (i.e., fair-coin flipping)
logspace probabilistic Turing machines with cut-point 1/2.

Finally, we mention that the above results may be extended to more general
space bounds by standard padding arguments, under the assumption that the
space bound $s$ is space-constructible and satisfies $s(n) = \Omega(\log n)$.
Assume that we have a space $O(s)$ uniform quantum circuit (defined
analogously to logspace uniform quantum circuits) acting on $s$ qubits, and
let $A$ be the language defined by the resulting circuits given some algebraic
cut-point $\beta$.
Define a new language
$\widetilde{A} = \left\{\left.x\,0\,1^{2^{s(|x|)}}\,\right|\,x\in A\right\}$.
It is straightforward to show that this language has logspace uniform
quantum circuits, and hence is in PL.
This follows from the fact that the suffix $0\,1^{2^{s(|x|)}}$ can easily be
recognized and ignored in logspace (following from the fact that $s$ is space
constructible, so we may simulate the machine that marks of $s(|x|)$ tape
squares and reject if this simulation requires more than logspace), after which
the original computation is performed on the prefix $x$ in space $s$, which is
logarithmic in the length of $x\,0\,1^{2^{s(|x|)}}$.
Given that $\widetilde{A}$ is in PL, it is also straightforward to show that
$A$ is in $\mbox{PrSPACE}(s)$ by similar arguments.
For a more thorough discussion of such techniques, see, e.g., Section 6.4
of~\cite{WagnerW86}.


\section{Proof of the main theorem}
\label{sec:proof_of_main}

In this section we present a proof of Theorem~\ref{thm:main}.
We begin by outlining the main ideas of the proof.
We then present various facts needed for the formal proof in Sections
\ref{sec:GapL} and \ref{sec:matrix}, and assemble the parts in Section
\ref{sec:assemble}.

The method used to prove the main theorem is similar to one used in a number
of other papers on space-bounded computation, particularly in
\cite{AllenderO96} and, in the quantum setting, \cite{Watrous99}; in short, the
long-term behavior of a given selective quantum process is determined by
performing various matrix operations on a matrix that determines the behavior
of the process.

Consider first the simpler case of stochastic processes described by Markov
chains.
Assume the Markov chain has state set $\{1,\ldots,N\}$, state 1 is the initial
state, and states $N-1$ and $N$ are absorbing states.
Assume the chain is described by a transition matrix $A$, and let $B$ be a
modification of $A$ where columns $N-1$ and $N$ are set to zero.
The probability that the process eventually enters state $N$ is given by
$\sum_t B^t[N,1]$.
Under the assumption that $B$ has eigenvalues strictly less than 1 in absolute
value (i.e., the process eventually enters state $N-1$ or $N$ with probability
1), the probability that the process eventually enters state $N$ is given by
\[
(I - B)^{-1}[N,1]\:=\:(-1)^{N+1}\frac{\op{det}((I-B)_{1,N})}{\op{det}(I-B)}.
\]
(We write $X_{i,j}$ to denote the matrix obtained by removing the
$i$th row and $j$th column of a given matrix $X$ throughout this paper.)
Thus, the problem of determining if this probability is strictly larger than
1/2 reduces to determining the sign of 
$(-1)^{N+1}\frac{\op{det}((I-B)_{1,N})}{\op{det}(I-B)} - \frac{1}{2}$.

In the situation that the chain above depends on some input string $x$, where
we assume $N$ is polynomial in $|x|$ and entries of $A$ are rational numbers
computable in logspace, it is possible to determine whether the chain
associated with $x$ enters state $N$ with probability exceeding 1/2 in PL as
follows: we define a GapL function that takes the same sign as
$(-1)^{N+1}\frac{\op{det}((I-B)_{1,N})}{\op{det}(I-B)} - \frac{1}{2}$, and
apply Theorem~\ref{thm:PL_GapL}.
The fact that such a GapL function exists follows from the properties of GapL
functions given by Theorems~\ref{thm:GapL_closure} and \ref{thm:determinant}.

Now, in the case of selective quantum processes defined by matrices of
algebraic numbers, the situation becomes somewhat more complicated, although
the main idea is the same.
The first issue we must face is the arithmetic with algebraic numbers.
We do not know how to approximate algebraic numbers to a sufficient degree of
accuracy for the above technique to work, supposing that this approximation
is to take place in deterministic logspace.
Instead, we approximate algebraic numbers by ratios of GapL functions;
Section~\ref{sec:GapL} below describes what we mean by a sufficiently
accurate approximation, and proves that this approximation can be achieved.
The second issue is that it is not immediate that a given selective quantum
process is governed by a single matrix as in the case of Markov chains.
Indeed this is the case, however, as we note in Section~\ref{sec:matrix}.
Here we also demonstrate how the basic technique from above can be extended
to the resulting matrices, which are not necessarily stochastic and may have
eigenvalues on the unit circle.


\subsection{GapL approximable numbers}
\label{sec:GapL}

We now define a class of numbers that can be efficiently approximated by
ratios of GapL functions, and then show that this class includes the algebraic
real numbers.

\begin{defn}
\label{def:G}
Let $\alpha\in\mathbb{R}$.
We say $\alpha$ is {\em GapL approximable} if there exist $f,g\in\mathrm{GapL}$
such that for all $n\geq 0$ we have $g(1^n)\not=0$ and
\[
\left|\frac{f(1^n)}{g(1^n)} - \alpha\right| < 2^{-n}.
\]
Denote the set of GapL approximable numbers by $\mathbb{G}$.
\end{defn}

\begin{theorem}
\label{thm:algebraic_in_G}
Let $\alpha$ be any real algebraic number.  Then $\alpha\in\mathbb{G}$.
\end{theorem}

\noindent
The proof of Theorem~\ref{thm:algebraic_in_G} relies on the following lemma.

\begin{lemma}
\label{lemma:bivariate}
Let $u_{0}$ and $u_{1}$ be bivariate integer polynomials and let $a_{0}$ and
$a_{1}$ be
integers.
Then there exists $f\in\mathrm{GapL}$ such that
\[
f(1^n,c) = \left\{
\begin{array}{ll}
u_{c}\left(f\left(1^{\lceil n/2\rceil},0\right),
f\left(1^{\lceil n/2\rceil},1\right)\right) & n \geq 2\\
a_{c} & n = 1
\end{array}
\right.
\]
for $n\geq 0$ and $c\in\{0,1\}$.
\end{lemma}
{\bf Proof.}
We will define a logspace NTM $M_{f}$ such that
$f(1^n,c) = \#M_{f}(1^n,c) - \#\overline{M_{f}}(1^n,c)$ satisfies the
recurrence in the statement of the lemma.

Write $u_{c}(X,Y) = \sum_{0\leq i,j\leq d}u_{c,i,j}X^{i}Y^{j}$.
To simplify the presentation of $M_{f}$, we define $M_{c}$ and $M_{c,i,j}$ for
$c\in\{0,1\}$, $0\leq i,j\leq d$ to be NTMs that take no input and satisfy
$\#M_{c} - \#\overline{M_{c}} = a_{c}$ and
$\#M_{c,i,j} - \#\overline{M_{c,i,j}} = u_{c,i,j}$.
The execution of $M_{f}$ may be described as follows:

\noindent\hrulefill\vspace{-2mm}
\begin{tabbing}
\rule{5mm}{0mm}\=\rule{5mm}{0mm}\=\rule{5mm}{0mm}\=\kill
Input: $(1^n,c)$.\\
Set $k=\lceil\log n\rceil$ and $s=1$.\\
Call P(c).\\
If $s=1$ then accept, otherwise reject.\\[3mm]
Procedure P(c)\\
\>If $k = 0$, simulate $M_{c}$ and set $s = -s$ if $M_{c}$ rejects.\\
\>Else\\
\>\>Guess $i,j\in\{0,\ldots,d\}$.\\
\>\>Simulate $M_{c,i,j}$ and set $s = -s$ if $M_{c,i,j}$ rejects.\\
\>\>Set $k=k-1$.\\
\>\>Repeat $i$ times: Call $P(0)$.\\
\>\>Repeat $j$ times: Call $P(1)$.\\
\>\>Set $k = k + 1$.\\
End Procedure P.\\[-8mm]
\end{tabbing}
\hrulefill\vspace{2mm}

\noindent
The variables $k$ and $s$ are ``global'', while $i$, $j$, and any auxiliary
variables needed by Procedure P are ``local''.
Since $i$, $j$, and all required auxiliary variables are constant in size,
$M_{f}$ will need to store only a constant amount of information for each
level of the recursion.
As the recursion will have depth at most logarithmic in $n$, $M_{f}$ requires
space $O(\log n)$ to implement the recursion.
Since each of the machines $M_{c}$ and $M_{c,i,j}$ require only constant
space, it follows that $M_{f}$ may be taken to run in space $O(\log n)$.

Now let us analyze the computation of $M_{f}$.
Each execution of Procedure P causes the computation of $M_{f}$ to branch
along several computation paths, each path having the effect of either leaving
$s$ unchanged or replacing $s$ with $-s$.
Let $r^{+}(k,c)$ denote the number of computation paths induced by calling
$P(c)$ for a given value of $k$ that leave $s$ unchanged, let $r^{-}(k,c)$
denote the number of computation paths induced by calling $P(c)$ that result
in $s$ being replaced by $-s$, and define $r(k,c) = r^{+}(k,c) - r^{-}(k,c)$.
Note that we have
$\#M_{f}(1^n,c) - \#\overline{M_{f}}(1^n,c) = r(\lceil\log n\rceil,c)$.
Since $\lceil\log\lceil n/2\rceil\rceil = \lceil\log n\rceil-1$ for any
integer $n\geq 2$, it remains to prove that $r$ obeys the recurrence
\[
r(k,c) = \left\{
\begin{array}{ll}
\displaystyle
u_{c}(r(k-1,0),r(k-1,1)) & k\geq 1\\[1mm]
a_{c} & k = 0.
\end{array}
\right.
\]
In case $k=0$, Procedure $P(c)$ induces $\#M_{c}$ computation paths that do
not modify $s$ and $\#\overline{M_{c}}$ paths that replace $s$ with $-s$, and
thus $r(0,c) = a_{c}$.
Now suppose $k\geq 1$ and assume that the number of paths induced by $P(b)$
that do not change $s$ (replace $s$ with $-s$) when $k$ is replaced by $k-1$
is described by $r^{+}(k-1,b)$ ($r^{-}(k-1,b)$, respectively) for each $b$.
For each pair $i,j$ that may be guessed, it may be proved (using the
binomial theorem) that the number of computation paths induced by the
remaining portion of $P(c)$ that have the effect of leaving $s$ unchanged
minus the number of paths that replace $s$ by $-s$ is given by
$u_{c,i,j}\,r(k-1,0)^i\,r(k-1,1)^j$.
We therefore conclude that
\begin{eqnarray*}
r(k,c) & = & \sum_{i,j}u_{c,i,j}\,r(k-1,0)^i\,r(k-1,1)^j\\
& = & u_{c}(r(k-1,0),r(k-1,1))
\end{eqnarray*}
for $k\geq 1$, which completes the proof.
\qed

\noindent {\bf Proof of Theorem~\ref{thm:algebraic_in_G}.}
Clearly we have $0\in\mathbb{G}$, so consider the case $\alpha\not=0$.
Let $p(x)=p_{d}x^{d}+\cdots+p_{0}$ be an integer polynomial such that
$p(\alpha)=0$, and assume without loss of generality that
$p^{\prime}(\alpha)\not=0$.

Lemma~\ref{lemma:bivariate} will allow us to use Newton's Method to approximate
$\alpha$ by GapL functions.
We have that there exist positive constants $\xi$ and $K$, where $\xi$ and
$\xi K$ are at most $1/2$, such that for $x_{0}\in(\alpha-\xi,\alpha+\xi)$ and
$x_{k+1} = x_{k} - p(x_{k})/p^{\prime}(x_{k})$
for $k\geq 0$, the inequality $|x_{k+1} - \alpha| \leq K|x_{k}-\alpha|^{2}$ is
satisfied for all $k\geq 0$.
Thus we have $|x_{k} - \alpha| < 2^{-2^k}$ for every $k\geq 0$.

Define
\begin{eqnarray*}
u_{0}(x,y) & = & \sum_{j=0}^{d}(j-1)p_{j}x^{j}y^{d-j}\\
u_{1}(x,y) & = & \sum_{j=1}^{d}j p_{j}x^{j-1}y^{d-j+1},
\end{eqnarray*}
and note that
\[
\frac{u_{0}(x,y)}{u_{1}(x,y)} = \frac{x}{y} - \frac{p(x/y)}{p^{\prime}(x/y)}.
\]
Let $a_{0},a_{1}\in\mathbb{Z}$, $a_{1}\not=0$, be such that
$|\alpha - a_{0}/a_{1}|<\xi$.
By Lemma~\ref{lemma:bivariate} there exists $f\in\mathrm{GapL}$ such that
\[
\frac{f(1^{n},0)}{f(1^{n},1)} \:=\:
\frac{u_{0}(f(1^{\lceil n/2\rceil},0),f(1^{\lceil n/2\rceil},1))}
{u_{1}(f(1^{\lceil n/2\rceil},0),f(1^{\lceil n/2\rceil},1))}
\:=\: \frac{f(1^{\lceil n/2\rceil},0)}{f(1^{\lceil n/2\rceil},1)}
- \frac{p\left(\frac{f(1^{\lceil n/2\rceil},0)}
{f(1^{\lceil n/2\rceil},1)}\right)}
{p^{\prime}\left(\frac{f(1^{\lceil n/2\rceil},0)}
{f(1^{\lceil n/2\rceil},1)}\right)}
\]
for $n\geq 2$, and $f(1,0)/f(1,1) = a_{0}/a_{1}$.
Consequently
\[
\left|\frac{f(1^n,0)}{f(1^n,1)} - \alpha\right| < 2^{-2^{\lceil\log n\rceil}}
\leq 2^{-n}
\]
for every $n\geq 1$.
We may now define $g,h\in\mathrm{GapL}$ that satisfy
\[
g(1^n) \:=\: \left\{
\begin{array}{ll}
f(1^n,0) & n\geq 1\\
a_{0} & n = 0
\end{array}
\right.
\]
and
\[
h(1^n) \:=\: \left\{
\begin{array}{ll}
f(1^n,1) & n\geq 1\\
a_{0} & n = 0,
\end{array}
\right.
\]
so that $|g(1^n)/h(1^n) - \alpha| < 2^{-n}$ for all $n\geq 0$.
Thus $\alpha\in\mathbb{G}$.
\qed

It is interesting to note that the set $\mathbb{G}$ is in fact a subfield
of the reals, following from a straightforward proof relying on the closure
properties of GapL functions.
It is also straightforward to prove that $\mathbb{G}$ contains some
transcendental numbers (such as $\pi$, for example) so $\mathbb{G}$ properly
contains the algebraic reals.


\subsection{Quantum processes and matrix problems}
\label{sec:matrix}

Next we prove that selective quantum processes may be described by transition
matrices in a manner similar to Markov chains.
It will simplify matters to note first that the selective quantum operations
and density matrices underlying a given selective quantum process may be
assumed to be real.

\begin{lemma}
\label{lemma:real_matrices}
Let $\{R_1,R_2,\ldots\}$ be a selective quantum process induced by selective
quantum operation $\mathcal{E} = \{A_{i,j}\}$ and initial state $\rho_{init}$.
Define real matrices $\{A_{i,j}^{\prime}\}$ and $\rho_{init}^{\prime}$ as
follows:
\[
\begin{array}{l}
A_{i,j}^{\prime}[2i^{\prime}-1,2j^{\prime}-1] \:=\: 
\Re(A_{i,j}[i^{\prime},j^{\prime}])\\
A_{i,j}^{\prime}[2i^{\prime}-1,2j^{\prime}] \:=\: 
\Im(A_{i,j}[i^{\prime},j^{\prime}])\\
A_{i,j}^{\prime}[2i^{\prime},2j^{\prime}-1] \:=\: 
-\Im(A_{i,j}[i^{\prime},j^{\prime}])\\
A_{i,j}^{\prime}[2i^{\prime},2j^{\prime}] \:=\: 
\Re(A_{i,j}[i^{\prime},j^{\prime}])
\end{array}
\]
and
\[
\begin{array}{l}
\rho_{init}^{\prime}[2i^{\prime}-1,2j^{\prime}-1] \:=\: 
\frac{1}{2}\,\Re(\rho_{init}[i^{\prime},j^{\prime}]) \\
\rho_{init}^{\prime}[2i^{\prime}-1,2j^{\prime}] \:=\: 
\frac{1}{2}\,\Im(\rho_{init}[i^{\prime},j^{\prime}]) \\
\rho_{init}^{\prime}[2i^{\prime},2j^{\prime}-1] \:=\: 
-\frac{1}{2}\,\Im(\rho_{init}[i^{\prime},j^{\prime}]) \\
\rho_{init}^{\prime}[2i^{\prime},2j^{\prime}] \:=\: 
\frac{1}{2}\,\Re(\rho_{init}[i^{\prime},j^{\prime}])
\end{array}
\]
Then $\{A_{i,j}^{\prime}\}$ and $\rho_{init}^{\prime}$ also induce the
selective quantum process $\{R_1,R_2,\ldots\}$.
\end{lemma}
The proof of this lemma is straightforward.

Recall that for $n\times n$ matrices $A$ and $B$, the Kronecker product
$A\otimes B$ is an $n^2\times n^2$ matrix satisfying
$(A\otimes B)[(i_0-1)n+i_1,(j_0-1)n+j_1]=A[i_0,j_0]\,B[i_1,j_1]$ for
$1\leq i_0,i_1,j_0,j_1\leq n$.
For fixed $n$, let us also define a mapping $\op{vec}$ from $n\times n$
matrices to $n^2$ dimensional (column) vectors as
$\op{vec}(A)[(i-1)n+j] = A[i,j]$ for $1\leq i,j\leq n$.
It is easy to prove that for $n\times n$ matrices $A$, $B$, and $C$ we have
$\op{vec}(ABC)=(A\otimes C^T)\op{vec}(B)$ and
$\op{tr}\left(A^T B\right)\:=\:\op{vec}(A)^T \op{vec}(B)$.

\begin{lemma}
\label{lemma:B_powers}
Let $\{A_{i,j}\,|\,0\leq i\leq m,\,1\leq j\leq l\}$ describe a selective
quantum operation, let $\rho_{init}$ be an initial state, and let
$\{R_1,R_2,\ldots\}$ be the induced selective quantum process.
For given $\beta$ define an $(n^2+2)\times(n^2+2)$ matrix $M$ as follows:
\[
M =
\left(
\begin{array}{ccc}
0 & 0 & 0 \\[2mm]
\op{vec}(\rho_{init}) & \sum_j A_{0,j}\otimes\overline{A_{0,j}} & 0\\[2mm]
-\beta & \op{vec}\left(\sum_j A_{1,j}^{\dagger}A_{1,j}\right)^{T} & 0
\end{array}
\right)
\]
Then all eigenvalues of $M$ are bounded in absolute value by 1.
Furthermore, for each nonnegative integer $t$ we have
$M^{t+2}[n^2+2,1] = \op{Pr}[R_1 = 0, \ldots, R_t = 0, R_{t+1} = 1]$.
\end{lemma}

\noindent
The following lemma is used in the proof of Lemma~\ref{lemma:B_powers}.
The proof is a modification of the proof of Lemma~1 in \cite{TerhalD98}.

\begin{lemma}
\label{lemma:eigenvalues}
Let $\{A_{i,j}\,|\,0\leq i\leq m,1\leq j\leq l\}$ satisfy
$\sum_{i,j}A_{i,j}^{\dagger}A_{i,j} = I$.
Then for each $i$, $\sum_j A_{i,j}\otimes \overline{A_{i,j}}$ has eigenvalues
bounded by 1 in absolute value.
\end{lemma}
{\bf Proof.}
Let $v\not=0$ and $\lambda$ satisfy
$(\sum_j A_{i,j}\otimes \overline{A_{i,j}})v=\lambda v$, and let $B$ be the
matrix such that $\op{vec}(B) = v$.
As $B\not=0$, there exists a unit vector $\ket{\psi}$ such that
$\bra{\psi}B\ket{\psi}\not=0$.
Define $C=\bra{\psi}B^{\dagger}\ket{\psi}B+\bra{\psi}B\ket{\psi}B^{\dagger}$
and write $|C|$ to denote $\sqrt{C^{\dagger}C}$.
Note that $|C|$ and $|C|+C$ are positive semidefinite, as $C$ is hermitian.
Let $F_i$ be as defined in Section~\ref{sec:quantum_processes}, so that
$F_i(B)=\lambda B$ and $F_i(B^{\dagger}) = \overline{\lambda}B^{\dagger}$.
Thus, we have
$\bra{\psi}F_i^k(|C|+C)\ket{\psi} = \bra{\psi}F_i^k(|C|)\ket{\psi} +
2|\bra{\psi}B\ket{\psi}|^2\Re(\lambda^k)$
for $k\geq 1$.
Since $|C|+C$ and $|C|$ are positive semidefinite, we have
$0\leq \bra{\psi}F_i^k(|C|+C)\ket{\psi} \leq \op{tr}(F_i^k(|C|+C))\leq
\op{tr}(|C|+C)$, and similarly
$0\leq\bra{\psi}F_i^k(|C|)\ket{\psi} \leq \op{tr}(|C|)$.
Consequently, $2|\bra{\psi}B\ket{\psi}|^2\Re(\lambda^k)$ is bounded
(independent of $k$).
As $|\bra{\psi}B\ket{\psi}|^2\not=0$, this implies $|\lambda|\leq 1$.
\qed

\noindent
{\bf Proof of Lemma~\ref{lemma:B_powers}.}
A straightforward computation shows the following:
\begin{eqnarray*}
B^{m+2}[n^2+2,1]
& = & \op{tr}\left(\sum_j A_{1,j} F^m_0(\rho_{init})A_{1,j}^{\dagger}\right)\\
& = & \op{tr}(F_1\circ F_0^m(\rho_{init}))\\
& = & \op{Pr}[R_1 = 0, \ldots, R_m = 0, R_{m+1} = 1].
\end{eqnarray*}
Thus it remains to show that all eigenvalues of $B$ are bounded by 1 in
absolute value.
By Lemma~\ref{lemma:eigenvalues} we have that all eigenvalues of
$\sum_j A_{0,j}\otimes\overline{A_{0,j}}$ are bounded by 1 in absolute value.
Since any nonzero eigenvalue of $B$ must be an eigenvalue of
$\sum_j A_{0,j}\otimes\overline{A_{0,j}}$, the required fact holds.
\qed

Lemmas \ref{lemma:real_matrices} and \ref{lemma:B_powers} will allow us to
translate the problem in Theorem~\ref{thm:main} regarding selective quantum
processes to an equivalent matrix problem.
The next theorem proves that this matrix problem is solvable in PL.

\begin{theorem}
\label{thm:matrix_problem}
Let $p$ be an integer polynomial satisfying
$p(n)\geq 2$ for $n\geq 0$, let
$\Omega = \{\alpha_1,\ldots,\alpha_k\}$ be any finite collection of real
algebraic numbers, and let $r_1,\ldots,r_k,s_1,\ldots,s_k\in\mathrm{GapL}$
such that each $s_l$ is nonzero on all inputs.
For each $x\in\Sigma^{\ast}$ let $M_x$ be a $p(|x|)\times p(|x|)$ matrix
defined as
\[
M_x[i,j] \:=\: \sum_{l=1}^k\frac{r_l(x,i,j)}{s_l(x,i,j)}\,\alpha_l,
\]
for $1\leq i,j\leq p(|x|)$.
Then if $M_x$ has eigenvalues bounded by 1 in absolute value and the series
$\sum_{t\geq 0}M_x^t[p(|x|),1]$ converges, we have
\[
\left\{x\in\Sigma^{\ast}\,\left|\sum_{t\geq 0}M_x^t[p(|x|),1] > 0\right.
\right\} \in PL.
\]
\end{theorem}
The proof of this theorem relies on a few technical facts, which we now state.
First, however, let us mention some notation: for any univariate polynomial
$f(X) = \sum_j f_jX^j$ or bivariate polynomial $f(X,Y) = \sum_{i,j} f_{i,j}
X^i Y^j$ we write $\|f\|$ to denote $\max_j\{|f_j|\}$ or
$\max_{i,j}\{|f_{i,j}|\}$, respectively.

\begin{lemma}
\label{lemma:rational_bound}
Let $u$ and $v$ be polynomials of degree at most $d\geq 1$ such that
$|v(0)|\geq\delta$.
Then for $|z| \leq \epsilon \leq \frac{\delta}{2\,d\,\|v\|}$, we have
$|v(z)| \geq \delta/2$ and
\[
\left|\frac{u(0)}{v(0)} - \frac{u(z)}{v(z)}\right| \:\leq\:
\frac{4\,\epsilon\,d\,\|u\|\,\|v\|}{\delta^{2}}.
\]
\end{lemma}
The proof is straightforward.

Let us now state a theorem due to Mahler that will be used below
(see pages 44--46 of \cite{Mahler61} for a proof).

\begin{theorem}[Mahler]
Let $f$ and $g$ be integer polynomials of degree $d_{f}$ and $d_{g}$,
respectively, and let $\alpha$ satisfy $f(\alpha)=0$ and $g(\alpha)\not=0$.
Then $|g(\alpha)|$ is greater than or equal to the quantity
\[
\frac{1}{(d_{f}\!+\!d_{g}\!-\!1)!\,\|f\|^{d_{g}}\,\|g\|^{d_{f}-1}\,
(|\alpha|^{d_{f}-1}\!+\!\cdots\!+\!|\alpha|+1)}.
\]
\label{thm:Mahler}
\end{theorem}

\begin{lemma}
\label{lemma:main_bound}
For any real algebraic number $\alpha$ there exist positive integer constants
$C_1$ and $C_2$ such that the following holds.
Let $g$ and $h$ be bivariate integer polynomials such that
$\|g\|,\|h\|\leq 2^N$ and $\op{deg}(g),\op{deg}(h)\leq N$, for $N\geq 2$, and
such that $\lim_{z\uparrow 1}\frac{g(\alpha,z)}{h(\alpha,z)}$ exists.
Then
\[
\left|\lim_{z\uparrow 1}\frac{g(\alpha,z)}{h(\alpha,z)} - 2^{-C_1 N^2}\right|
\geq 2^{-C_1 N^2}.
\]
Furthermore, for $\xi$ and $\widetilde{\alpha}$ satisfying
$\xi\leq 2^{-C_2 N^2}$ and
\mbox{$|\alpha-\widetilde{\alpha}|\leq\xi^{2N+1}$}, we
have $h(\widetilde{\alpha},1-\xi)\not=0$ and
\[
\left|\lim_{z\uparrow 1}\frac{g(\alpha,z)}{h(\alpha,z)} -
\frac{g(\widetilde{\alpha},1-\xi)}{h(\widetilde{\alpha},1-\xi)}\right|
< 2^{-C_1 N^2}.
\]
\end{lemma}
{\bf Proof.}
First, we note that by Lemma~\ref{lemma:rational_bound} there exist positive
constants $C_3$ and $C_4$, depending only on $\alpha$, such that for any
polynomials $p$ and $q$ satisfying $\op{deg}(p),\op{deg}(q)\leq N$,
$\|p\|,\|q\|\leq 2^{2N} (1+|\alpha|)^N$, and $|q(0)|\geq\delta$ for
$\delta>0$, we have $|q(\nu)|\geq\delta/2$ and
\[
\left|\frac{p(0)}{q(0)} - \frac{p(\nu)}{q(\nu)}\right| \leq
\frac{|\nu|\,2^{C_4 N^2}}{\delta^2}
\]
whenever $|\nu|\leq\delta\,2^{-C_3 N^2}$.
(Of course these inequalities are not tight---rather they are chosen to
simplify arithmetic and notation below.)
Furthermore, by Theorem~\ref{thm:Mahler} there exists a positive constant
$C_5$, again depending only on $\alpha$, such that for any polynomial $p$
satisfying $\op{deg}(p)\leq N$ and $\|p\|\leq 2^{2N}$ we have
$|p(\alpha)|\geq 2^{-C_5 N^2}$ whenever $p(\alpha)\not=0$.
Without loss of generality assume $C_5\geq 2+2|\alpha|$.

Now, define $u(x) = g(\alpha,1-x)$ and $v(x) = h(\alpha,1-x)$.
We may write
\[
u(x) = \sum_{j=0}^N a_j(\alpha)x^j\;\;\;\;\;\mbox{and}\;\;\;\;\;
v(x) = \sum_{j=0}^N b_j(\alpha)x^j
\]
for integer polynomials $a_j$ and $b_j$, 
$0\leq j\leq N$, satisfying
$\op{deg}(a_j),\op{deg}(b_j)\leq N$ and $\|a_j\|,\|b_j\|\leq 2^{2N}$.
Let $k = \op{min}\{j\,|\,b_j(\alpha)\not=0\}$.
As $b_k(\alpha)\not=0$, we have $|b_k(\alpha)|\geq 2^{-C_5 N^2}$.
Similarly, $|a_k(\alpha)| \geq 2^{-C_5 N^2}$ in case $a_k(\alpha)$ is nonzero.
Define $u_0(x) = u(x)/x^k$ and $v_0(x) = v(x)/x^k$.
As we assume $\lim_{z\uparrow 1}\frac{g(\alpha,z)}{h(\alpha,z)}$ exists, we
must have $a_j(\alpha)=0$ for $j<k$, and hence $u_0$ and $v_0$ are polynomials.
Furthermore, we have $\lim_{z\uparrow 1}\frac{g(\alpha,z)}{h(\alpha,z)}
=\frac{u_0(0)}{v_0(0)}=\frac{a_k(\alpha)}{b_k(\alpha)}$ and
$\frac{u(x)}{v(x)} = \frac{u_0(x)}{v_0(x)}$ whenever $v(x)\not=0$.
Consequently
\begin{equation}
\left|\lim_{z\uparrow 1}\frac{g(\alpha,z)}{h(\alpha,z)}\right|
\:\geq\: \frac{2^{-C_5 N^2}}{2^{2N}\,(1+|\alpha|)^N}
\:\geq\: 2^{-2\,C_5 N^2}
\label{eq:otherbound}
\end{equation}
whenever the limit is nonzero.
Also note the following: $\op{deg}(u_0),\op{deg}(v_0)\leq N$,
$\|u_0\|,\|v_0\|\leq 2^{2N}(1+|\alpha|)^{N}$, and
$|v_0(0)| = |b_k(\alpha)| \geq 2^{-C_5 N^2}$.
Thus, for $\xi \leq 2^{-(C_3+C_5) N^2}$ it follows that
$|v_0(\xi)| \geq \frac{1}{2}\,2^{-C_5 N^2}$ and
\begin{equation}
\left|\frac{u_0(0)}{v_0(0)} - \frac{u_0(\xi)}{v_0(\xi)}\right|
\leq \xi\,2^{(C_4+2C_5) N^2}.
\label{eq:xi_bound1}
\end{equation}

Now assume $\xi \leq 2^{-(C_3+C_5) N^2}$ is fixed, and define
$r(x) = g(\alpha - x,1-\xi)$ and $s(x) = h(\alpha - x,1-\xi)$.
We have $\op{deg}(r),\op{deg}(s)\leq N$,
$\|r\|,\|s\|\leq 2^{2N}(1+|\alpha|)^N$, and
\[
|s(0)| \:=\: |v(\xi)| \:\geq\: \frac{1}{2}\xi^{N}2^{-C_5 N^2}.
\]
Thus, for
$|\alpha-\widetilde{\alpha}|\leq\xi^{2N+1}\leq\frac{1}{2}\,\xi^N
2^{-(C_3+C_5) N^2}$ we conclude that
$s(\alpha-\widetilde{\alpha})=h(\widetilde{\alpha},1-\xi)\not=0$ and
\begin{equation}
\left|\frac{r(0)}{s(0)} - \frac{r(\alpha - \widetilde{\alpha})}
{s(\alpha-\widetilde{\alpha})}\right| \:\leq\:
4\,\xi\,2^{(C_4+2 C_5) N^2}.
\label{eq:xi_bound2}
\end{equation}
By (\ref{eq:xi_bound1}) and (\ref{eq:xi_bound2}) we therefore have
\begin{eqnarray}
\left|\,\lim_{z\uparrow 1}\frac{g(\alpha,z)}{h(\alpha,z)} -
\frac{g(\widetilde{\alpha},1-\xi)}{h(\widetilde{\alpha},1-\xi)}\right|
& \leq & \left|\frac{u_0(0)}{v_0(0)} - \frac{u_0(\xi)}{v_0(\xi)}\right|
+ \left|\frac{r(0)}{s(0)} - \frac{r(\alpha - \widetilde{\alpha})}
{s(\alpha - \widetilde{\alpha})}\right|\nonumber\\
& < & 5\,\xi\,2^{(C_4+2 C_5) N^2}.
\label{eq:anotherbound}
\end{eqnarray}

Now, define $C_1=\lceil 2\,C_5+1\rceil$ and
$C_2=C_1 + \lceil C_3+C_4+2\,C_5+3\rceil$.
By (\ref{eq:otherbound}) we have
\[
\left|\lim_{z\uparrow 1}\frac{g(\alpha,z)}{h(\alpha,z)}-
2^{-C_1 N^2}\right| \geq 2^{-C_1 N^2},
\]
as $2^{-C_1 N^2} \leq \frac{1}{2}2^{-2C_5 N^2}$.
Furthermore, for $\xi$ and $\widetilde{\alpha}$ satisfying
$\xi\leq 2^{-C_2 N^2}$ and $|\alpha - \widetilde{\alpha}|\leq \xi^{2N+1}$, we
have $\xi\leq 2^{-(C_3+C_5)N^2}$, and thus by (\ref{eq:anotherbound}) we have
\[
\left|\,\lim_{z\uparrow 1}\frac{g(\alpha,z)}{h(\alpha,z)} -
\frac{g(\widetilde{\alpha},1-\xi)}{h(\widetilde{\alpha},1-\xi)}\right|
\:<\: 5\cdot 2^{-C_2 N^2}\,2^{(C_4+2 C_5) N^2}\:<\:2^{-C_1 N^2}
\]
as required.
\qed

\noindent
{\bf Proof of Theorem~\ref{thm:matrix_problem}.}
First, we outline briefly the main idea of the proof.
(Throughout the proof we let $p$ denote $p(|x|)$, as $|x|$ is always the point
at which $p$ is evaluated).
Under the assumption that $M_x$ has eigenvalues bounded in absolute value by
1 and the series $\sum_{t\geq 0}M_x^t[p,1]$ converges, we have
\begin{equation}
\sum_{t\geq 0}M_x^t[p,1]\:=\:
\lim_{z\uparrow 1}\frac{\op{det}((I - z M_x)_{1,p})}{\op{det}(I - z M_x)}.
\label{eq:thing_to_approximate}
\end{equation}
We approximate the algebraic numbers comprising $M_x$ by ratios of GapL
functions, and we approximate the limit by substituting for $z$ a quantity
very close to 1.
Relying on the fact that the determinants of matrices defined by GapL
functions are also in GapL (Theorem~\ref{thm:determinant}),  our
approximation of (\ref{eq:thing_to_approximate}) will be a ratio of GapL
functions.
Based on these GapL functions, together with a bound on the error of the
approximation, we define a GapL function $F$ such that $F(x)>0$ if and only if
$\sum_{t\geq 0}M_x^t[p,1]>0$.
By Theorem~\ref{thm:PL_GapL}, this suffices to prove the theorem.
In the remainder of the proof, we define the function $F$ and demonstrate that
it is indeed the case that $F(x)>0$ if and only if
$\sum_{t\geq 0}M_x^t[p,1]>0$.
For convenience we assume $|x|\geq 2$, since $F$ may be modified on inputs of
length 0 and 1 without changing the fact that it is a GapL function

Let $\alpha$ be a real algebraic number such that
$\mathbb{Q}[\alpha]=\mathbb{Q}[\alpha_1,\ldots,\alpha_k]$; such an $\alpha$
always exists as $\mathbb{Q}[\alpha_1,\ldots,\alpha_k]$ is a finite degree
(separable) extension of $\mathbb{Q}$ (see, e.g., \cite{Isaacs94}, page 284).
Let $d$ be the degree of the minimal polynomial of $\alpha$ and fix positive
integers $m$ and $B$ and integer polynomials $q_1,\ldots,q_k$ so that
$\alpha_l = \frac{q_l(\alpha)}{m}$, $\op{deg}(q_l)\leq d$, and
$\|q_l\|\leq B$ for $1\leq l\leq k$.
Define bivariate integer polynomials $w_1,\ldots,w_k$ as
$w_l(y_1,y_2) = y_2^d\,q_l(y_1/y_2)$, and note that
$\op{deg}(w_l)\leq d$ and $\|w_l\|\leq B$ for $1\leq l\leq k$.
Also note that $d$, $m$, $B$, $q_1,\ldots,q_k$ and $w_1,\ldots,w_k$ depend
only on $\Omega$, and not on the input~$x$.

Define
\[
h(x) \: = \: m\,\prod_{i=1}^p\prod_{j=1}^p\prod_{l=1}^k\,s_l(x,i,j).
\]
By Theorem~\ref{thm:GapL_closure}, $h\in\mathrm{GapL}$.
Let $E_x(y)$ be a $p\times p$ matrix defined by
\[
E_x(y)[i,j] \: = \: h(x)\sum_{l=1}^k\frac{q_l(y)\,r_l(x,i,j)}{m\,s_l(x,i,j)}.
\]
For each $i,j$, $E_x(y)[i,j]$ is an integer polynomial in $y$.
We have $E_x(\alpha) = h(x)M_x$.
Next, define
\begin{eqnarray*}
u_x(y,z) \!\!\! & = & \!\!\!
(-1)^{1+p}\,h(x)\,\op{det}((h(x)\,I-z\,E_x(y))_{1,p}),\\
v_x(y,z) \!\!\! & = & \!\!\! \op{det}(h(x)\,I-z\,E_x(y)).
\end{eqnarray*}
Note that there exists a positive integer constant $C$ such that
$\op{deg}(u_x),\op{deg}(v_x)\leq |x|^C$ and\linebreak
$\|u_x\|,\|v_x\|\leq 2^{|x|^C}$.
Given that $M_x$ has eigenvalues bounded by 1 in absolute value and
$\sum_{t\geq 0}M_x^t[p,1]$ converges for each $x$, we have
\[
\lim_{z\uparrow 1}\frac{u_x(\alpha,z)}{v_x(\alpha,z)}
\:=\: \sum_{t\geq 0}M_{x}^t [p,1].
\]
By Lemma~\ref{lemma:main_bound}, there exist positive integer constants $C_1$
and $C_2$ such that
$v_x\left(\widetilde{\alpha},1-2^{-C_2 |x|^{2C}}\right)\not=0$,
\begin{equation}
\left|\,\sum_{t\geq 0}M_x^t[p,1]-2^{-C_1 |x|^{2C}}\,\right|\:\geq\:
2^{-C_1 |x|^{2C}},
\label{eq:inequality1}
\end{equation}
and
\begin{equation}
\left|\,\sum_{t\geq 0}M_x^t[p,1] - \frac{u_x\left(\widetilde{\alpha},
1-2^{-C_2 |x|^{2C}}\right)}{v_x\left(\widetilde{\alpha},1-2^{-C_2 |x|^{2C}}
\right)}\,\right| \:<\: 2^{-C_1 |x|^{2C}}
\label{eq:inequality2}
\end{equation}
whenever $|\widetilde{\alpha}-\alpha|<2^{-3\,C_2 |x|^{3C}}$.

By Theorem~\ref{thm:algebraic_in_G} there exist GapL functions $f$ and $g$
such that $\left|f(1^n)/g(1^n)-\alpha\right| < 2^{-n}$ for each $n\geq 0$.
Define $\nu(|x|) = 3\,C_2 |x|^{3C}$ and write
$\widetilde{\alpha} = \frac{f(1^{\nu(|x|)})}{g(1^{\nu(|x|)})}$.
The value $\widetilde{\alpha}$ will be our approximation of $\alpha$.
Also define $\mu(|x|) = C_2 |x|^{2C}$.
The value $1-2^{-\mu(|x|)}$ will be substituted for $z$ in order to
approximate the limit.

Next, define
\[
a(x,i,j)
=\sum_{l=1}^k\left[\left(\prod_{i^{\prime}=1}^p\prod_{j^{\prime}=1}^p
\prod_{l^{\prime}=1}^k [(i^{\prime},j^{\prime},l^{\prime})\not=(i,j,l)]
s_{l^{\prime}}(x,i^{\prime},j^{\prime})\right)
r_l(x,i,j)\,w_l\left(f(1^{\nu(|x|)}),g(1^{\nu(|x|)})\right)\right]
\]
for $1\leq i,j\leq p$, and let $A_x$ denote the $p\times p$ matrix defined
by $A_x[i,j] = a(x,i,j)$.
Here we let $[(i^{\prime},j^{\prime},l^{\prime})\not=(i,j,l)]$ denote the
value 1 or 0 depending on whether
$(i^{\prime},j^{\prime},l^{\prime})\not=(i,j,l)$ or
$(i^{\prime},j^{\prime},l^{\prime})=(i,j,l)$, respectively.
By Theorem~\ref{thm:GapL_closure}, $a\in\mathrm{GapL}$.
Note that $A_x = (g(1^{\nu(|x|)}))^d E_x(\widetilde{\alpha})$.
Define
\[
b(x,i,j) \:=\: h(x)(g(1^{\nu(|x|)}))^d\,2^{\mu(|x|)}[i=j]
-(2^{\mu(|x|)}-1)a(x,i,j),
\]
and let $B_x$ be the $p\times p$ matrix defined by $B_x[i,j] = b(x,i,j)$.
Thus, we have
\[
B_x \:=\: h(x)(g(1^{\nu(|x|)}))^d 2^{\mu(|x|)}I-(2^{\mu(|x|)}-1)A_x.
\]
By Theorem~\ref{thm:GapL_closure}, $b\in\mathrm{GapL}$.
Next, define
\begin{eqnarray*}
U(x)\!\!\! & = &\!\!\! (-1)^{p+1} h(x) (g(1^{\nu(|x|)}))^d 2^{\mu(|x|)}
\op{det}((B_x)_{1,p}),\\
V(x) \!\!\!& = &\!\!\! \op{det}(B_x).
\end{eqnarray*}
By Theorem~\ref{thm:GapL_closure} and Theorem~\ref{thm:determinant},
$U,V\in\mathrm{GapL}$.
Finally, define
\[
F(x) \:=\: 2^{C_1 |x|^{2C}}U(x)V(x) - (V(x))^2.
\]
By Theorem~\ref{thm:GapL_closure}, $F\in\mathrm{GapL}$.

It remains to show that for every $x\in\Sigma^{\ast}$, $F(x)>0$ if and only if
$\sum_{t\geq 0}M_x^t[p,1]>0$.
By the above, it may be verified that
\[
U(x) \:=\: (g(1^{\nu(|x|)}))^{dp}\,
2^{p\mu(|x|)}\, u_x(\widetilde{\alpha},1-2^{\mu(|x|)})
\]
and
\[
V(x) \:=\: (g(1^{\nu(|x|)}))^{dp}\,
2^{p\mu(|x|)}\, v_x(\widetilde{\alpha},1-2^{\mu(|x|)}).
\]
Thus we have $V(x)\not=0$.
Furthermore $F(x)>0$ if and only if
\[
\frac{U(x)}{V(x)}> 2^{-C_1 |x|^{2C}},
\]
which is equivalent to
\begin{equation}
\frac{u_x\left(\widetilde{\alpha},1-2^{\mu(|x|)}\right)}
{v_x\left(\widetilde{\alpha},1-2^{\mu(|x|)}\right)}\:>\:2^{-C_1 |x|^{2C}}.
\label{eq:lastone}
\end{equation}
By (\ref{eq:inequality1}) and (\ref{eq:inequality2}), the inequality in
(\ref{eq:lastone}) holds if and only if
$\sum_{t\geq 0}M_x^t[p,1]>0$, as required.
\qed


\subsection{Completion of the proof}
\label{sec:assemble}

Now Theorem~\ref{thm:main} follows in straightforward fashion.
Let $p$, $q$, $\{\alpha_1,\ldots,\alpha_k,\beta\}$, $a_l$, $b_l$, $c_l$,
and $d_l$, $1\leq l\leq k$, and $\{R_{x,1},R_{x,2},\ldots\}$ be as in the
statement of the theorem.
By Lemma~\ref{lemma:real_matrices} we may assume $\{\alpha_1,\ldots,\alpha_k\}$
are real algebraic numbers, since otherwise we modify the $a_l$, $b_l$,
$c_l$, and $d_l$ functions and take the real and imaginary parts of
$\{\alpha_1,\ldots,\alpha_k\}$ accordingly.
Let $M_x$ be the $(p(|x|)^2+2)\times (p(|x|)^2+2)$ matrix described in
Lemma~\ref{lemma:B_powers} for each input $x$.
Using Theorem~\ref{thm:GapL_closure} it is routine to define GapL functions
$r_l$ and $s_l$, $1\leq l\leq k+1$, such that
\[
M_x[i,j] \:=\: \sum_{l=1}^{k+1}\frac{r_l(x,i,j)}{s_l(x,i,j)}\,\alpha_l,
\]
where we write $\alpha_{k+1}=\beta$.
By Lemma~\ref{lemma:B_powers} we see that the series
$\sum_{t\geq 0}M_x^t[p(|x|)^2 + 2,1]$ must converge, since the sum is $-\beta$
plus a sum over probabilities of mutually exclusive events.
Furthermore, we have
\[
\op{Pr}[\exists\,t:\,R_{x,1}=\cdots=R_{x,t-1}=0,\,R_{x,t}=1]>\beta
\]
if and only if $\sum_{t\geq 0}M_x^t[p(|x|)^2+2,1]>0$.
Theorem~\ref{thm:main} now follows from Theorem~\ref{thm:matrix_problem}.


\subsection*{Acknowledgments}

I would like to thank Eric Allender for a helpful discussion on GapL functions,
and Marcus Schaefer and Pradyut Shah for their suggestions regarding the
proof of Theorem~\ref{thm:algebraic_in_G}.


\bibliographystyle{plain}


\end{document}